\documentclass[fleqn,12pt]{article}
\usepackage{epsfig}
\begin{document}
\textheight 22cm
\textwidth 15cm
\noindent
{\Large \bf Zonal flow generation in ion temperature gradient mode turbulence}
\newline
\newline
J.Anderson\footnote{elfja@elmagn.chalmers.se}, H. Nordman, R. Singh, J. Weiland
\newline
Department of Electromagnetics, EURATOM-VR Association
\newline
Chalmers University of Technology, G\"{o}teborg, Sweden
\newline
\newline
\begin{abstract}
\noindent
In the present work the zonal flow (ZF) growth rate in toroidal ion-temperature-gradient (ITG) mode turbulence including the effects of elongation is studied analytically. The scaling of the ZF growth with plasma parameters is examined for typical tokamak parameter values.
The physical model used for the toroidal ITG driven mode is based on the ion continuity and ion temperature equations whereas the ZF evolution is described by the vorticity equation. The results indicate that a large ZF growth is found close to marginal stability and for peaked density profiles and these effects may be enhanced by elongation.
\end{abstract}
\newpage
\renewcommand{\thesection}{\Roman{section}}
\section{Introduction}
\indent
The important role played by plasma flows for the reduction of energy transport in tokamak regimes of enhanced confinement is now widely acknowledged. Of particular interest for the improved confinement regimes are the nonlinearly self-generated zonal flows~\cite{a11}. These are radially localized and strongly sheared flows in the poloidal direction. In experiments, zonal flow levels above the neoclassical prediction has been observed in connection with transport barriers~\cite{a12} in the edge (H-mode barrier) and core (internal transport barriers) regions. Due to the strong shear stabilization of the driving instabilities associated with these flows, they are crucial for the dynamical self-regulation and saturation of the underlying turbulence and anomalous transport fluxes~\cite{a13}. Theoretically, the generation and evolution of zonal flows due to drift mode turbulence has been extensively studied in recent years, both analytically ~\cite{a13} -~\cite{a47} and in computer simulations using gyrokinetic ~\cite{a22} -~\cite{a24} and advanced fluid models ~\cite{a25} -~\cite{a27}. In particular, the gyrokinetic simulations of ITG mode turbulence reported in the Cyclone work Ref. ~\cite{a9} indicate that there is a strong excitation of zonal flows close to marginal stability where the non-linearly generated flows were able to damp out the turbulence resulting in a non-linear up-shift in the critical temperature gradient needed to obtain transport for longer time scales.

In the present paper, the excitation of zonal flows by toroidal ion-temperature gradient (ITG) turbulence is studied analytically. A system of equations is derived which describes the coupling between the background ITG turbulence and the zonal flow modes. The model used is an advanced fluid model for ITG modes~\cite{a28} including effects of elongated flux surfaces. Previous work using this model indicates that the ITG growth is rather insensitive to the shape of the flux surface~\cite{a29} -~\cite{a30}. The analytical technique used here closely follow Ref.~\cite{a16} where the coupling between the zonal flow modes, driven by Reynolds stress forces, and the ITG turbulence is described by a kinetic equation for wave packets. The present work extends the previous study by studying effects of non circular geometry on the zonal flow growth rate. The result is an analytical expression for the growth rate of the zonal flow instability in elongated equilibria.

The scaling of the zonal flow growth rate with plasma parameters is investigated. In particular, the role of plasma elongation on the generation of zonal flow is examined. A resonance in the zonal flow excitation level is found close to marginal stability, consistent with the Cyclone work~\cite{a9}. For peaked density profiles (small $\epsilon_n = 2 L_n/L_B$), there is a substantially increased excitation of zonal flow when the elongation is increased whereas for most other cases the effects of elongation are weak. Moreover, the zonal flow excitation grows linearly with the wavenumber for zero collisional damping whereas for non-zero damping the zonal flow excitation is significantly reduced.

The paper is organized as follows. In Section II the physical model for the toroidal ITG modes is presented. The equations describing the coupling between the background ITG turbulence and the zonal flow modes is presented in Section III. Section IV is dedicated to the results and a discussion thereof. Finally there is a summary in section V.
\section{Toroidal ion-temperature-gradient driven modes}
The description used for toroidal ITG driven modes consists of the ion continuity and ion temperature equations. The effect of parallel ion motion is weak in the reactive model and is hence neglected magnetic shear can, however, modify the non-linear up shift as found in Ref.~\cite{a21}. For simplicity, effects of electron trapping and finite beta effects are neglected in this work. The ion-temperature and ion-continuity equations can be written
\begin{eqnarray}
\frac{\partial n_{i}}{\partial t} +\nabla \cdot \left( n_{i} \vec{v}_{E} + n_{i} \vec{v}_{\star i} \right) + \nabla \cdot \left( n_{i} \vec{v}_{pi} + n_{i} \vec{v}_{\pi i} \right) & = & 0 \\ 
 \frac{3}{2} n_{i} \frac{d T_{i}}{dt} + n_{i} T_{i} \nabla \cdot \vec{v}_{i} + \nabla \cdot \vec{q}_{i} & = & 0.
\end{eqnarray}
Here $\vec{v}_E$ is the $\vec{E} \times \vec{B}$ velocity, $\vec{v}_{\star}$ is the diamagnetic drift velocity, $\vec{v}_{pi}$ is the polarization drift velocity and $\vec{v}_{\pi i}$ is the stress tensor drift velocity and $\vec{q}_i$ is the ion heat flux. The derivative is defined as $d/dt = \partial / \partial t + \rho_s c_s \vec{z} \times \nabla \tilde{\phi} \cdot \nabla$ and $n$, $\phi$, $T_i$ are the ion density, the electrostatic potential and the ion temperature, respectively. With the additional definitions $\tilde{n} = \delta n / n_0$, $\tilde{\phi} = e \delta \phi /T_e$, $\tilde{T}_i = \delta T_i / T_{i0}$ as the normalized ion particle density, the electrostatic potential and the ion temperature, respectively. In the forthcoming equations $\tau = T_i/T_e$, $\vec{v}_{\star} = \rho_s c_s \vec{y}/L_n $, $\rho_s = c_s/\Omega_{ci}$ where $c_s=\sqrt{T_e/m_i}$, $\Omega_{ci} = eB/m_i c$. We also define $L_f = - \left( d ln f / dr\right)^{-1}$, $\eta_i = L_n / L_{T_i}$, $\epsilon_n = 2 L_n / R$ where $R$ is the major radius and $\alpha_i = \tau \left( 1 + \eta_i\right)$. The perturbed variables are normalized with the additional definitions $\tilde{n} = L_n/\rho_s \delta n / n_0$, $\tilde{\phi} = L_n/\rho_s e \delta \phi /T_e$, $\tilde{T}_i = L_n/\rho_s \delta T_i / T_{i0}$ as the normalized ion particle density, the electrostatic potential and the ion temperature, respectively. The perpendicular length scale and time are normalized to $\rho_s$ and $L_n/c_s$, respectively. The geometrical quantities are calculated in the strong ballooning limit ($\theta = 0 $, $g\left(\theta = 0, \kappa \right) = 1/\kappa$ (Ref.~\cite{a29}) where $g\left( \theta \right)$ is defined by $\omega_D \left( \theta \right) = \omega_{\star} \epsilon_n g\left(\theta \right)$). The equations 1 and 2 can now be simplified to
\begin{eqnarray}
\frac{\partial \tilde{n}}{\partial t} - \left(\frac{\partial}{\partial t} - \alpha_i \frac{\partial}{\partial y}\right)\nabla^2_{\perp} \tilde{\phi} + \frac{\partial \tilde{\phi}}{\partial y} - \epsilon_n g \frac{\partial}{\partial y} \left(\tilde{\phi} + \tau \left(\tilde{n} + \tilde{T}_i \right) \right) = \nonumber \\
- \left[\phi,n \right] + \left[\phi, \nabla^2_{\perp} \phi \right] + \tau \left[\phi, \nabla^2_{\perp} \left( n + T_i\right) \right] \\
\frac{\partial \tilde{T}_i}{\partial t} - \frac{5}{3} \tau \epsilon_n g \frac{\partial \tilde{T}_i}{\partial y} + \left( \eta_i - \frac{2}{3}\right)\frac{\partial \tilde{\phi}}{\partial y} - \frac{2}{3} \frac{\partial \tilde{n}}{\partial t} = \nonumber \\
- \left[\phi,T_i \right] + \frac{2}{3} \left[\phi,n \right].
\end{eqnarray}
Here $\left[ A ,B \right] = \partial A/\partial x \partial B/\partial y - \partial A/\partial y \partial B/\partial x$ is the Poisson bracket. Linearizing the Equations 3 and 4 and using Boltzmann distributed electrons ($\tilde{n}_i = \tilde{n}_e = \tilde{\phi}$) gives
\begin{eqnarray}
\left(\omega - \left(1 - \left(1 + \tau\right) \epsilon_n g \right) k_y + \left( \omega + \alpha_i k_y\right) k_{\perp}^2\right)\tilde{\phi} + \tau \epsilon_n g k_y \tilde{T}_i = 0 \\
\left(\omega + \frac{5}{3} \tau \epsilon_n g k_y \right) \tilde{T}_i - \left( \left(\eta_i - \frac{2}{3}\right) k_y + \frac{2}{3} \omega \right)\tilde{\phi} = 0
\end{eqnarray}
The corresponding dispersion relation and solutions are
\begin{eqnarray}
0 & = & \omega^2 \left( 1 + k_{\perp}^2 \right) - \omega k_y \left( 1 - \left(1 + \frac{10\tau}{3} \right) \epsilon_n g - k_{\perp}^2 \left(\alpha_i + \frac{5}{3} \tau \epsilon_n g \right)\right) \\
& + & \tau \epsilon_n g k_y^2 \left(\eta_i - \frac{7}{3} + \frac{5}{3} \left(1 + \tau \right) \epsilon_n g + \frac{5}{3} \alpha_i k_{\perp}^2\right) \\
\omega_r & = & \frac{k_y}{2\left( 1 + k_{\perp}^2\right)} \left( 1 - \left(1 + \frac{10\tau}{3} \right) \epsilon_n g - k_{\perp}^2 \left( \alpha_i + \frac{5}{3} \tau \epsilon_n g \right)\right)  \\
\gamma & = & \frac{k_y}{1 + k_{\perp}^2} \sqrt{\tau \epsilon_n g\left( \eta_i - \eta_{i th}\right)}
\end{eqnarray}
where $\omega = \omega_r + i \gamma$ and 
\begin{eqnarray}
\eta_{i th} \approx \frac{2}{3} - \frac{1}{2 \tau} + \frac{1}{4 \tau \epsilon_n g} + \epsilon_n g\left( \frac{1}{4 \tau} + \frac{10}{9 \tau}\right).
\end{eqnarray}
FLR effects in the $\eta_{i th}$ are neglected, however, they are important when the group velocities are to be determined. The group velocities ($v_{gj} = \partial \omega_r/\partial k_j$) are in the long wavelength limit ($k^2_{\perp} << 1$) given by,
\begin{eqnarray}
v_{gx} & = & - k_x k_y \left(1 + \left( 1 + \eta_i\right) \tau - \left(1 + \frac{5 \tau}{3} \right) \epsilon_n g \right) \\
v_{gy} & = & \frac{1}{2} \left( 1 - \left( 1 + \frac{10 \tau}{3}\right) \epsilon_n g\right).
\end{eqnarray}
The instability of the mean flow in the linear regime sets in when the velocity of the mean flow modulations are close to the group velocities of the small scale wave packet. 
\noindent
In Figure 1 the effects of elongation and $\epsilon_n$ on the $\eta_i$-mode stability is illustrated. The growth rate (normalized the the electron diamagnetic drift frequency) as a function of $\eta_i$ with elongation ($\kappa$) and $\epsilon_n$ as parameters is displayed. The results are shown for $\tau = 1$ and $\epsilon_n = 0.2$; $\kappa=1.0$ (diamonds) and $\kappa = 2.0$ (squares) and  $\epsilon_n = 2.0$; $\kappa = 1.0$ (+ curve) and $\kappa = 2.0$ (* curve).
\begin{figure}
  \includegraphics[height=1.0\textheight]{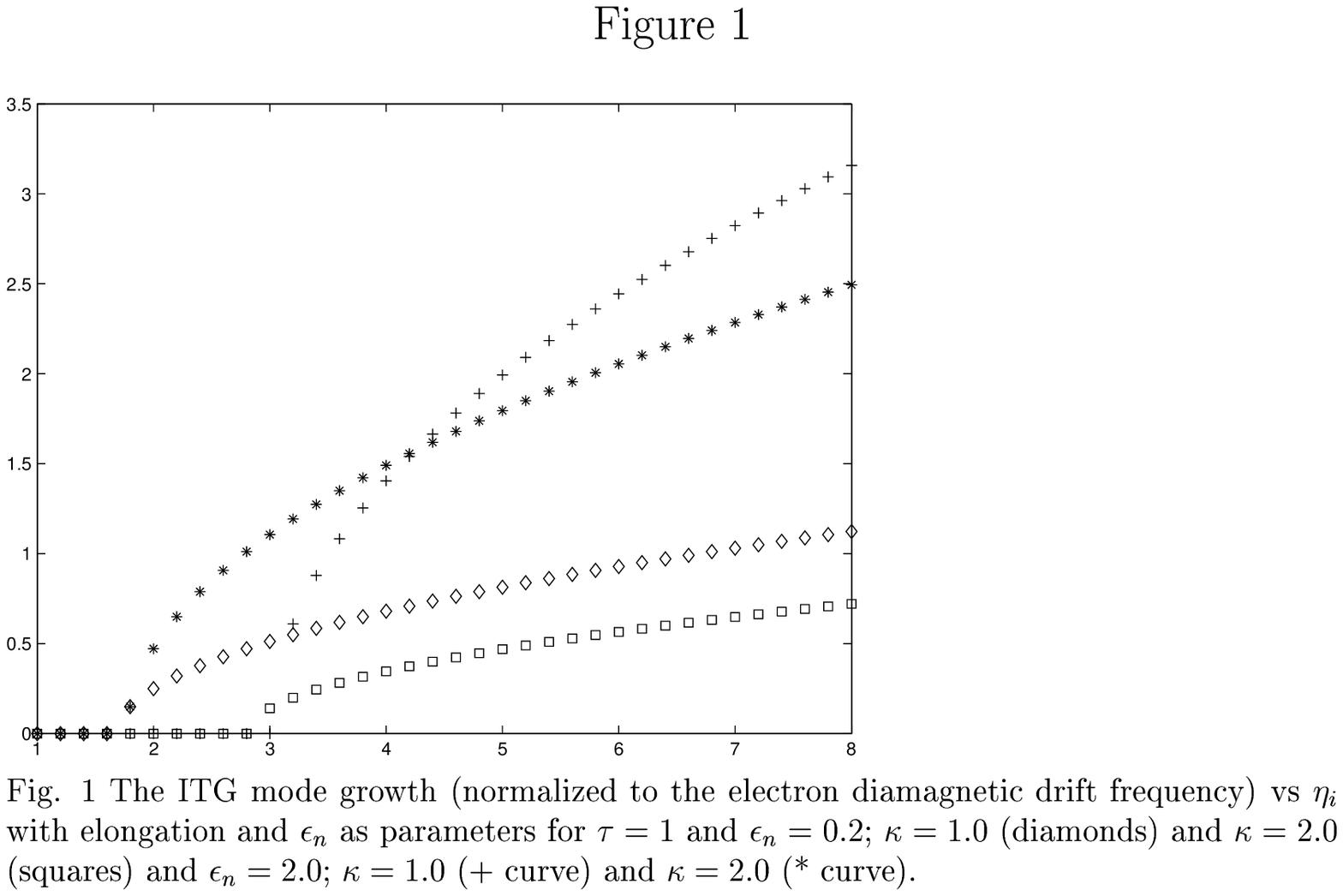}
\end{figure}
\section{The wave kinetic equation and zonal flow evolution}
In this section the problem of how to construct the adiabatic invariant in ITG driven turbulence and deriving the evolution equations for zonal flows and ITG perturbations is summarized. The method has been described in detail in Ref.~\cite{a16} (and References therein) and only a brief summary is given here. An alternative statistical approach, resulting in a modified wave kinetic equation (see Eq. 26 below), is presented in Ref.~\cite{a47} which also contains an extensive discussion of and comparison with the approach used here. In describing the large scale plasma flow dynamics it is assumed that there is a sufficient spectral gap between the small scale fluctuations and the large scale flow. The electrostatic potential is represented as a sum of fluctuating and mean quantities
\begin{eqnarray}
\phi(X,x,T,t) = \Phi(X,T) + \tilde{\phi}(x,t)
\end{eqnarray}
where $\Phi(X,T)$ is the mean flow potential. The coordinates $\left( X, T\right)$, $\left( x,t \right)$ are the spatial and time coordinates for the mean flows and small scale fluctuations, respectively. From equations 3 and 4 we get after neglecting the FLR non-linearities, 
\begin{eqnarray}
\frac{\partial \tilde{\phi}}{\partial t} + \left(1 - \left( 1 + \tau \right) \epsilon_n g\right)\frac{\partial \tilde{\phi}}{\partial y} - \tau \epsilon_n g \frac{\partial \tilde{T}_i}{\partial y} & = & - \left[\Phi, \tilde{n}\right] \\
\frac{\partial \tilde{T}_i }{\partial t} - \frac{7}{3} \tau \epsilon_n g \frac{\partial \tilde{T}_i}{\partial y} + \left( \eta_i - \frac{2}{3} \left(1 + \tau\right) \epsilon_n g \right) \frac{\partial \tilde{\phi}}{\partial y} & = & -\left[\Phi, \tilde{T}_i\right].
\end{eqnarray}
Here, the interaction between the ITG perturbations have been omitted(see discussion after Eq. 26 for this.) In order to determine the generalized wave action density $N_k = |\Psi_k|^2$ we introduce the normal coordinates  $\Psi_k = \tilde{\phi}_k + \alpha_k \tilde{T}_i$, where $\alpha_k$ is to be calculated. Multiplying equation 16 by $\alpha_k$ and adding it to equation 15 gives
\begin{eqnarray}
\frac{\partial }{\partial t} \left(\tilde{\phi}_k + \alpha_k \tilde{T}_{ik} \right) & + & \left(1 - \left(1 + \tau \right) \epsilon_n g + \alpha_k \left( \eta_i - \frac{2}{3} \left( 1 + \tau\right) \epsilon_n g \right) \right)\frac{\partial \tilde{\phi_k}}{\partial y} \nonumber \\ 
& - & \left( \frac{7}{3} \tau \epsilon_n g \alpha_k + \tau \epsilon_n g \right) \frac{\partial\tilde{T}_{ik}}{\partial y} = - \left[\Phi,\tilde{\phi}_k + \alpha_k \tilde{T}_{ik} \right]
\end{eqnarray}
The normal coordinates are found if the equation is rewritten as in Ref. ~\cite{a16}
\begin{eqnarray}
\frac{\partial \Psi_k }{\partial t} + V_k \frac{\partial \Psi_k }{\partial y} = - \left[\Phi, \Psi_k \right]
\end{eqnarray}
where 
\begin{eqnarray}
V_k & = & 1 - \left(1 + \tau \right) \epsilon_n g + \alpha_k \left( \eta_i - \frac{2}{3} \left( 1 + \tau\right) \epsilon_n g \right) \\
\alpha_k & = & - \frac{\frac{7 \tau}{3}\epsilon_n g \alpha_k + \tau \epsilon_n g }{V_k}
\end{eqnarray}
which gives 
\begin{eqnarray}
\alpha_k = \frac{-\frac{1}{2} \left( 1 - \epsilon_n g + \frac{4 \tau}{3} \epsilon_n g\right) + i \sqrt{\epsilon_n g \left(\eta_i - \eta_{i th}\right)}}{\eta_i - \frac{2}{3} \left(1 + \tau \right) \epsilon_n g }
\end{eqnarray}
The linear relations between $\tilde{\phi}_k$ and $\tilde{T}_{ik}$ enables one to express $\Psi_k$ and $N_k$ as
\begin{eqnarray}
\Psi_k & = & \tilde{\phi}_k + \alpha_k \tilde{T}_{ik} = \frac{2 i \gamma_k}{\Delta_k + i\gamma_k } \tilde{\phi}_k \\
N_k & = & |\Psi_k|^2 = \frac{4 \gamma_k^2}{\Delta_k^2 + \gamma_k^2}|\tilde{\phi}_k|^2 \\ 
\Delta_k & = & \frac{k_y}{2}\left( 1 - \epsilon_n g + \frac{4 \tau}{3} \epsilon_n g\right) \\
\gamma_k & = & k_y \sqrt{\epsilon_n g \left(\eta_i - \eta_{i th}\right)}
\end{eqnarray}
The Eqs 22-23 and 25 describe the normal variables, the adiabatic invariant found from the normal variables and the linear ITG growth rate, respectively. The wave kinetic equation see Refs.~\cite{a16},~\cite{a2} -~\cite{a8} for the generalized wave action $N_k$ in the presence of mean plasma flow due to the interaction between mean flow small scale fluctuations is
\begin{eqnarray}
\frac{\partial }{\partial t} N_k(x,t) & + & \frac{\partial }{\partial k_x} \left( \omega_k + \vec{k} \cdot \vec{v}_g \right)\frac{\partial N_k(x,t)}{\partial x} - \frac{\partial }{\partial x} \left( \vec{k} \cdot\vec{v}_g\right) \frac{\partial N_k(x,t)}{\partial k_x} \nonumber \\
& = &  \gamma_k N_k(x,t) - \Delta\omega N_k(x,t)^2
\end{eqnarray}
In this analysis it is assumed that the RHS is approximately zero (stationary turbulence). The role of non-linear interactions among the ITG fluctuations (here represented by a non-linear frequency shift $\Delta\omega$) is to balance linear growth rate. In the case when $ \gamma_k N_k(x,t) - \Delta\omega N_k(x,t)^2 = 0$, the expansion of equation 26 is made under the assumption of small deviations from the equilibrium spectrum function; $N_k = N_k^0 + \tilde{N}_k$ where $\tilde{N}_k$ evolves at the zonal flow time and space scale  $\left( \Omega, q_x, q_y = 0\right)$, as
\begin{eqnarray}
- i \left(\Omega - q_x v_{gx} + i \gamma_k \right) \tilde{N}_k = k_y \frac{\partial^2 }{\partial x^2} \Phi \frac{\partial N_k^0 }{\partial k_x}
\end{eqnarray}
\begin{eqnarray} 
\tilde{N}_k = - q_x^2 k_y \frac{\partial N_k^0}{\partial k_x} \frac{i}{\Omega - q_x v_{gx} + i \gamma_k}
\end{eqnarray}
The evolution equations for the zonal flows is obtained after averaging the ion-continuity equation over the magnetic flux surface and over fast scales. including a damping term ~\cite{a4}. The average of the ion continuity equation over the magnetic surface and over fast small scales employing the quasi-neutrality the evolution of the mean flow is obtained
\begin{eqnarray}
\frac{\partial}{\partial t} \nabla_x^2 \Phi -\mu \nabla_x^4 \Phi = \left(1 + \tau \right) \nabla_x^2 \left<\frac{\partial}{\partial x} \tilde{\phi}_k \frac{\partial}{\partial y}\tilde{\phi}_k \right> + \tau \nabla_x^2 \left<\frac{\partial}{\partial x} \tilde{\phi}_k \frac{\partial}{\partial y}\tilde{T}_{ik} \right>
\end{eqnarray}
where it is assumed that only the small scale self interactions are the important interactions in the RHS~\cite{a31}. Using typical tokamak parameters ($T_i = T_e = 10 kev$, $n_i = n_e = 10^{20}m^{-3}$, $r = 1m$, $R = 3m$) $\mu = 0.78 \nu_{ii} \sqrt(r/R)$ and $\nu_{ii} = 10^{-12} n_i/T_i^{3/2}$ and $\nu_{ii}$ is the ion-ion collision frequency, $T_i$ is the ion temperature in electron volts. Using typical tokamak parameters it is found that $\mu \approx 50$.
Expressing the Reynolds stress terms in Eq. 29 in $N_k$ we obtain
\begin{eqnarray}
\left(-i \Omega - \mu q_x^2 \right) \Phi = \left(1 + \tau  + \tau \delta \right) \int d^2 k k_x k_y |\tilde{\phi}_k|^2
\end{eqnarray}
where $\delta$ is a $k$ independent factor
\begin{eqnarray}
\delta = \frac{\Delta_k k_y}{\Delta_k^2 + \gamma_k^2} \left(\eta_i - \frac{2}{3} \left(1 + \tau \right)  \epsilon_n g\right).
\end{eqnarray}
Utilize equations 23, 28 and 30 gives,
\begin{eqnarray}
\left(-i \Omega - \mu q_x^2 \right) = - i q_x^2 \left(1 + \tau  + \tau \delta \right) \frac{\Delta_k^2 + \gamma_k^2}{4 \gamma_k^2} \nonumber \\
\int d^2 k k_y^2 k_x \frac{\partial N_k^0}{\partial k_x} \frac{q_x v_{gx}}{\Omega - q_x v_{gx} + i \gamma_k}.
\end{eqnarray}
where $k_x k_y$ is substituted using the group velocity as 
\begin{eqnarray}
-k_x k_y = \frac{v_{gx}}{1 + \left(1 + \eta_i \right) \tau  - \left(1 +\frac{5 \tau}{3} \right) \epsilon_n}. \nonumber
\end{eqnarray}
It is now assumed that the short scale turbulence is close to marginal state (or stationary state, $\gamma_k$ is small). Integrating by parts in $k_x$ and assuming a monochromatic wave packet $N_k^0 = N_0 \delta\left(k - k_0\right)$ gives
\begin{eqnarray}
\left(\Omega + i \mu q_x^2 \right)\left(\Omega - q_x v_{gx}\right)^2 = - q_x^2 \left(1 + \tau  + \tau \delta \right) \frac{\Delta_k^2 + \gamma_k^2}{4 \gamma_k^2} k_y^2 N_0 \Omega 
\end{eqnarray}
In the special case of $\mu = 0$, the third order dispersion relation for zonal flow $\Omega$ reduces to
\begin{eqnarray}
\Omega = q_x v_{gx} + i q_x k_y\sqrt{\left(1 + \tau  + \tau \delta \right) \frac{\Delta_k^2 + \gamma_k^2}{4 \gamma_k^2} N_0}.
\end{eqnarray}
In expressing the zonal flow growth in dimensional form making use of the relation $\left(\Delta^2_k + \gamma_k^2\right)/\left( 4 \gamma_k^2 \right) N_0 = |\tilde{\phi}|^2$ it is assumed that the mode coupling saturation level is reached~\cite{a40} 
\begin{eqnarray}
\tilde{\phi} = \frac{\gamma}{\omega_{\star}}\frac{1}{k_y L_n}
\end{eqnarray}
\section{Results and discussion}
An algebraic equation Eq. 33 describing the zonal flow growth rate including the effects of elongation is derived where the 3rd order dispersion relation is solved numerically and the zonal flow growth rates found are compared to the linear toroidal ITG growth rates. 

In Figure 2 the zonal flow growth rate (normalized to the ITG growth rate) as a function of $\epsilon_n (= 2 L_n/L_B $) is displayed with elongation ($\kappa$) as parameter for $\eta_i = 4$, $\tau=1$, $q_x = 0.3$, $\mu = 0$. The results are shown for $\kappa = 1$ (+ curve) and $\kappa = 2$ (* curve). We note that a strong excitation of zonal flows (with $\gamma/\gamma_{ITG} \geq 1$) is obtained for small $\epsilon_n$ where a resonance is found. Similar results have been reported earlier in analytical calculations~\cite{a21} and in the Cyclone non-linear (collisionless, electrostatic) numerical calculations of ITG mode turbulence~\cite{a24}. In the numerical calculations there were a non-linear up-shift in the linear threshold from $R/L_T \approx 4$ to $R/L_T \approx 6$ and the other parameters were $\epsilon_n = 0.9$, $\tau=1$. The zonal flow growth is increased with increased elongation for small $\epsilon_n$.
\begin{figure}
  \includegraphics[height=1.0\textheight]{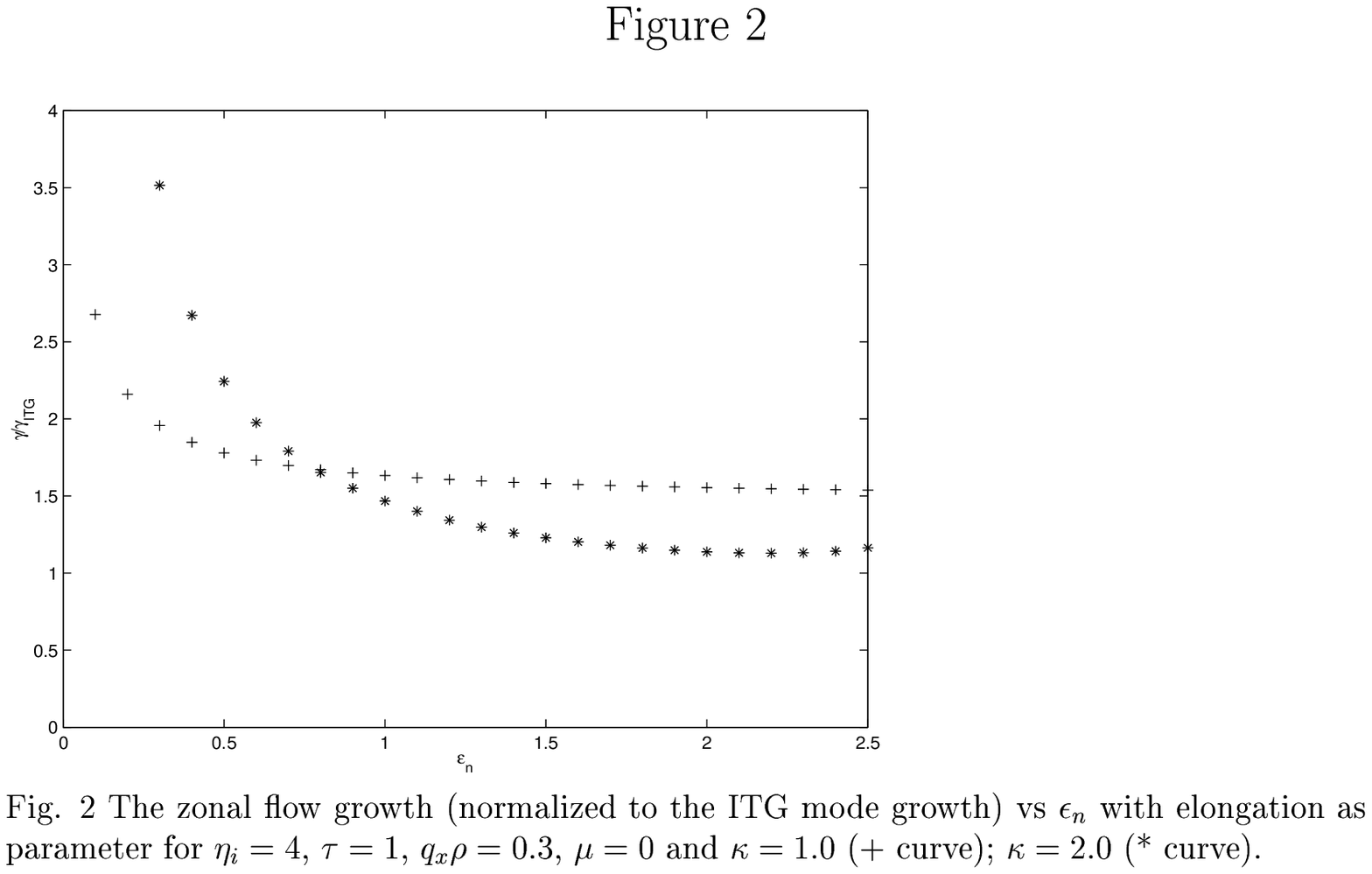}
\end{figure}

Next, the effects of elongation and $\epsilon_n$ on the zonal flow growth rate is studied. Figure 3 shows the zonal flow growth rate (normalized to the ITG growth rate) as a function of $\kappa$ with $\epsilon_n$ as parameter and $\eta_i = 4$ and the other parameters as in Fig. 2. For peaked density profiles the zonal flow growth rate is increased with elongation and  $\gamma/\gamma_{ITG} \geq 1$ is obtained whereas for flat density profiles (large $\epsilon_n$) the effects of elongation on the zonal flow growth are rather weak.
\begin{figure}
  \includegraphics[height=1.0\textheight]{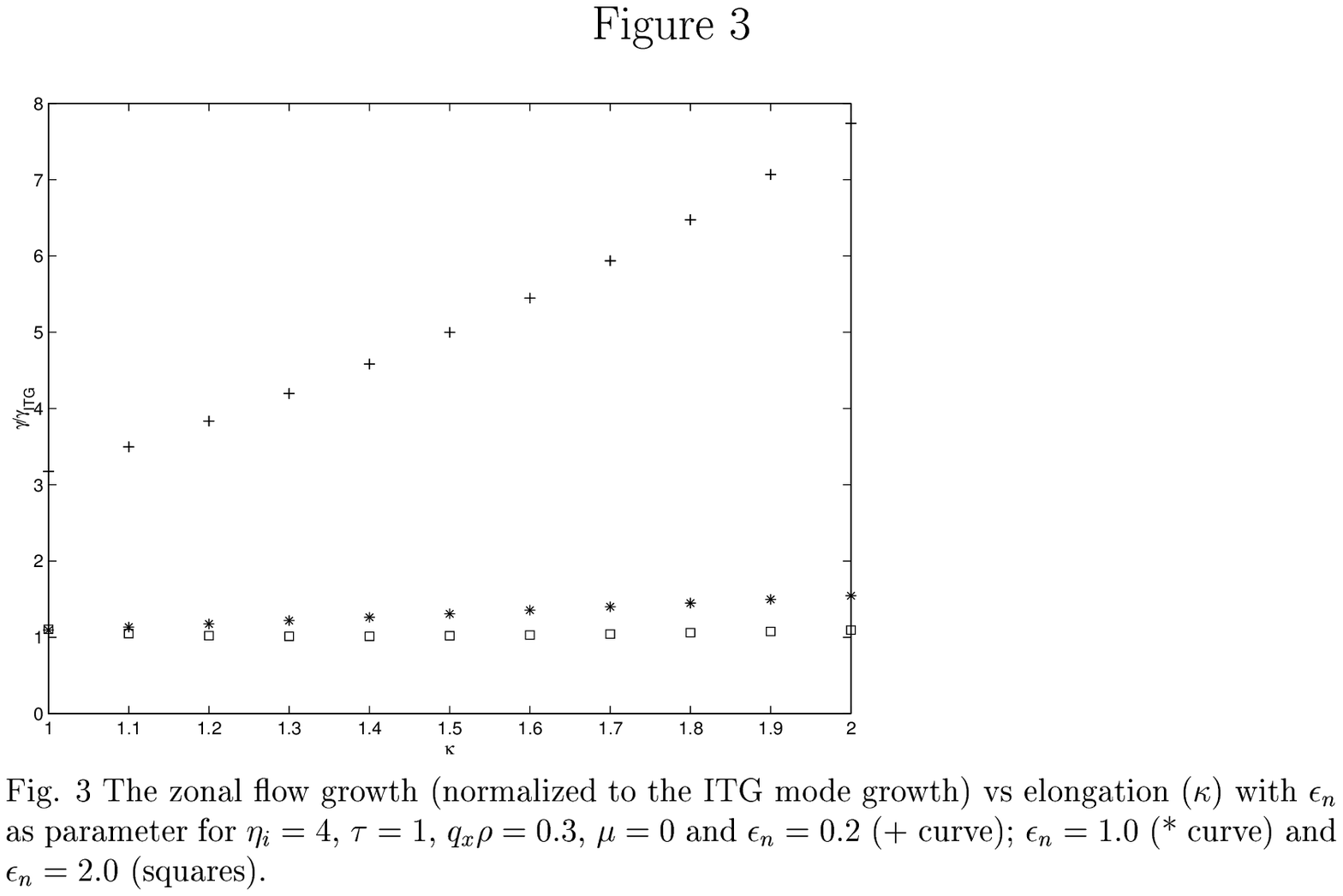}
\end{figure}

Figure 4 illustrates the effect of collisional damping ($\mu$) and elongation on the zonal flow growth (normalized to the ITG growth rate). The normalized zonal flow growth rate is shown as a function of $\mu$ with $\kappa$ as a parameter. The other parameters are as in Figure 3 with $\kappa=1$ (* curve) and $\kappa=2$ (+ curve). The zonal flow growth rate is decreased as the damping is increased and the effects of elongation are weak in this parameter regime. For $\mu = 50$ (typical value) a reduction of zonal flow growth with approximately 50 \% is obtained.
\begin{figure}
  \includegraphics[height=1.0\textheight]{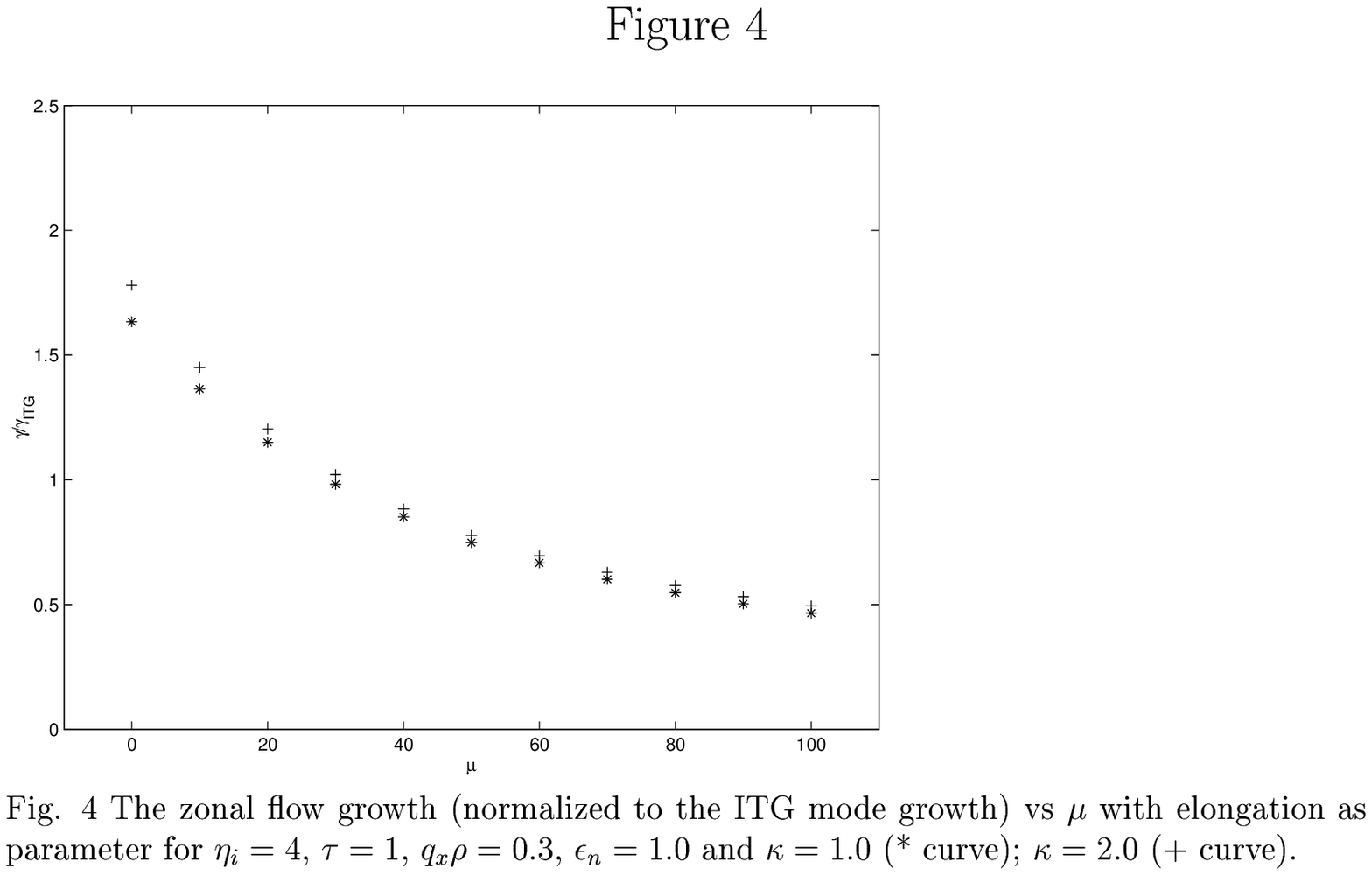}
\end{figure}

In Figure 5 the zonal flow growth rate as a function of the zonal flow wave number $q_x$ with elongation and damping as parameters is displayed. The other parameters are as in Fig. 3 with $\mu = 0$; $\kappa = 1$ (* curve) and $\kappa = 2$ (diamonds) whereas $\mu = 50$; $\kappa=1$ (+ curve) and $\kappa=2$ (squares). The effect of a non-zero damping is strong on the zonal flow growth rate and for zero damping the growth rate is linearly dependent on $q_x$. Moreover, the effects of elongation are small.
\begin{figure}
  \includegraphics[height=1.0\textheight]{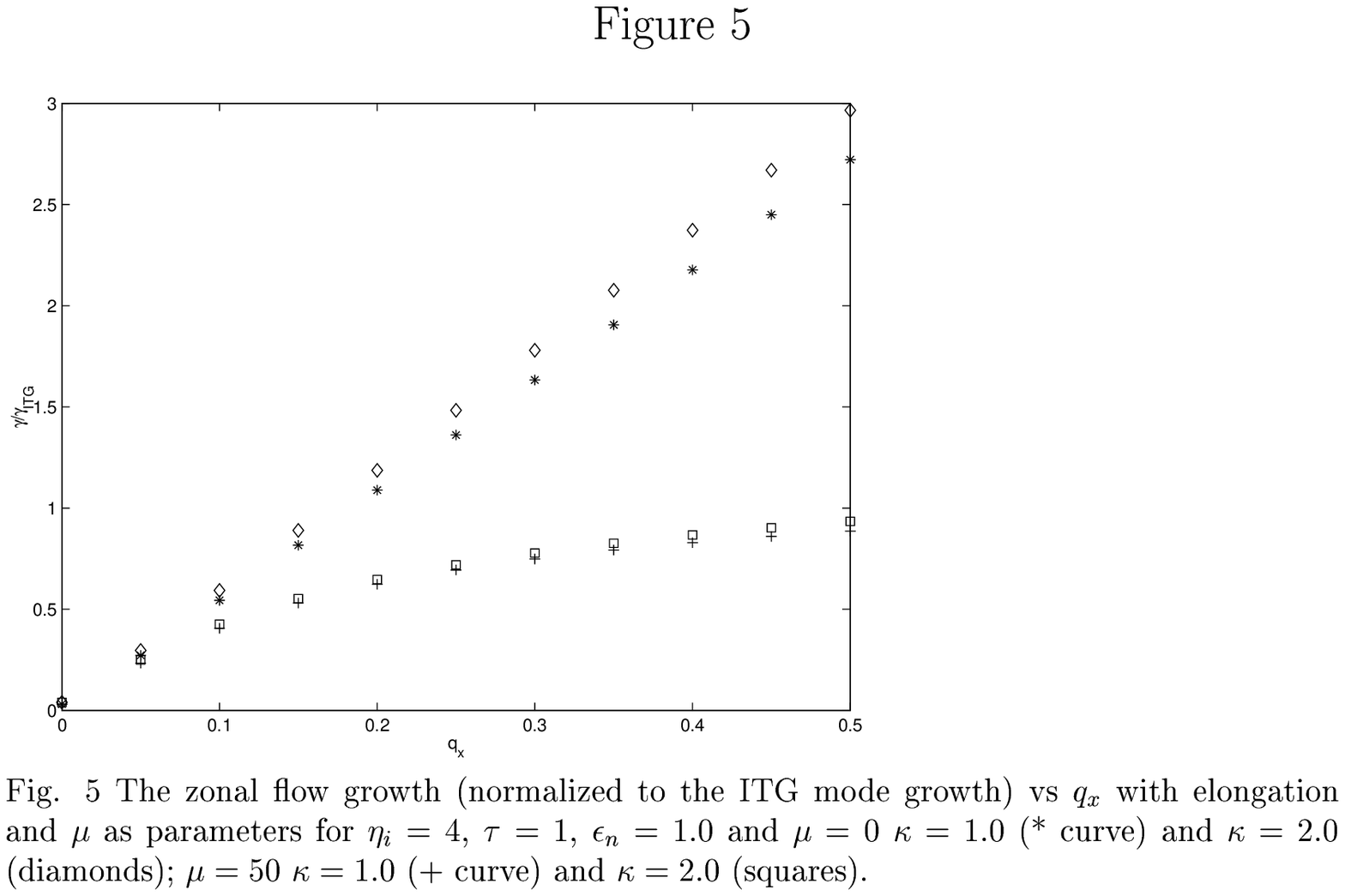}
\end{figure}

\section{Summary}
\noindent
An algebraic equation which describes the zonal flow growth rate in the presence of toroidal ITG turbulence is derived and solved numerically in the strong ballooning limit. The model for the ITG driven mode is based on the ion-continuity and the ion-temperature equations. The present model is electrostatic and effects of trapping are neglected. The evolution of zonal flows is described by the vorticity equation and the time evolution of the adiabatic invariant in toroidal ITG turbulence is determined by the wave kinetic equation. This gives a system of equations that couples the zonal flow and the ITG driven mode perturbations. The equilibrium model used includes the effects of elongated flux surfaces. 

A strong generation of zonal flows is obtained for peaked density profiles. The effects of elongation on the generation of zonal flows for realistic tokamak parameters are rather weak, however, a resonance of the zonal flow generation is found close to marginal stability as in Ref.~\cite{a21} which is consistent with the results reported in Ref.~\cite{a24}. When damping is included in the model there is a significant reduction in the zonal flow generation for increasing zonal flow radial wave numbers whereas for zero damping the zonal flow generation is linearly dependent on the radial wavenumber.

In short, this work indicates that for most parameter regimes, the effects of elongation on zonal flows driven by pure ITG modes are rather weak. However, the total effect on transport levels and turbulence may be significant since there is a resonance in the generation of zonal flows and this effect may be enhanced by elongation. For peaked density profiles, a strong excitation of zonal flows is found with $\gamma/\gamma_{ITG} \ge 1$. 


\newpage
\begin{thebibliography}{200}
\bibitem{a11} A. Hasegawa, C. G. Mcclennan and Y. Kodama, Phys. Fluids {\bf 22}, 2122 (1979) 
\bibitem{a12} R. E. Bell, F. M. Levinton, S. H. Batha {\it et al.}, Phys. Rev. Lett.{\bf 81}, 1429 (1998)
\bibitem{a13} H. Biglari, P.H. Diamond and P.W. Terry, Phys. Fluids B {\bf 2}, 1 (1990)
\bibitem{a14} P. H. Diamond and Y. B. Kim, Phys. Fluids B, {\bf 5} 2343 (1991)
\bibitem{a16} A. I. Smolyakov, P. H. Diamond, M. V. Medvedev Phys. Plasmas, {\bf 7} 3987 (2000)
\bibitem{a17} L. Chen, Z. Lin and R. White, Phys. Plasmas {\bf 7}, 3129 (2000)
\bibitem{a18} A.I. Smolyakov, P. H. Diamond and M. Malkov, Phys. Rev. Lett. {\bf 84}, 491 (2000)
\bibitem{a19} M.A. Melkov, P.H. Diamond and A.I. Smolyakov, Phys. Plasmas {\bf 8}, 1553 (2001)
\bibitem{a20} P.N.Guzdar, R.G.Kleva and L. Chen, Phys. Plasmas {\bf 8}, 459 (2001)
\bibitem{a21} S. Mahajan, J. Weiland Plasma Phys. Contr. Fusion, {\bf 42} 987 (2000)
\bibitem{a47} J. A. Krommes, C.-B. Kim, Phys. Rev. E {\bf 62}, 8508 (2000) 
\bibitem{a22} Z. Lin, T.S Hahm, W.W. Lee, W.M. Tang and R.B. White, Science 281, 1835 (1998)
\bibitem{a23} Z. Lin, T.S Hahm, W.W. Lee, W.M. Tang and P.H. Diamond, Phys. Rev. Lett. {\bf 83}, 3645 (1999)
\bibitem{a24} A. Dimits, T.J. Williams, J.A. Byers and B.I. Cohen, Phys. Rev. Lett. {\bf 77},{\bf 71} (1996)
\bibitem{a25} G. Hammett, M. Beer, W. Dorland, S.C. Cowley and S.A. Smith, Plasma Phys. Controlled Fusion {\bf 35}, 937 (1993)
\bibitem{a26} R.E. Waltz, G.D. Kerbel and A.J. Milovich, Phys. Plasmas{\bf 1}, 2229 (1994)
\bibitem{a27} M. A. Beer, Ph.D. dissertation, Princeton Univ., 1995
\bibitem{a9} A. Dimits, G. Bateman, M. A. Beer et al. Phys. Plasmas, {\bf 7} 969 (2000)
\bibitem{a28} J. Weiland, Collective Modes in Inhomogeneous Plasmas, Kinetic and Advanced Fluid Theory (IOP Publishing Bristol 2000) 115
\bibitem{a32} J. Anderson, H. Nordman, J. Weiland, Plasma Phys. Controlled Fusion {\bf 42}, 545 (2000)
\bibitem{a29} J. Anderson, H. Nordman, J. Weiland, Phys. Plasmas, {\bf 8} 180 (2001)
\bibitem{a30} T Rafiq, J. Anderson, M. Nadeem, M. Persson Plasma Phys. Contr. Fusion, {\bf 43} 1363 (2001) 
\bibitem{a2} A. A. Vedenov, A. V. Gordeev, L. I. Rudakov Plasma Phys.,{\bf 9}, 719 (1967)
\bibitem{a8} A. I. Smolyakov, P. H. Diamond Phys. Plasmas, {\bf 6} 4410 (1999)
\bibitem{a4} P. H. Diamond, Y. B. Kim Phys. Fluids B, {\bf 3} 1626 (1991)
\bibitem{a31} W. Horton, D. Choi, P. Terry Phys. Fluids, {\bf 23} 590 (1980)
\bibitem{a5} P. H. Diamond, S. Champaux, M. Malkov {\it et al.} Nucl. Fusion, {\bf 41} 1067 (2001) 
\bibitem{a7} S. Mahajan, J. Weiland Varenna Proc. 281 (2000)
\bibitem{a40} H. Nordman and J. Weiland Nucl. Fusion {\bf 29}, 251 (1989)
\end{thebibliography}
\end{document}